\documentclass[aps,prl,twocolumn]{revtex4}

\usepackage{graphicx}
\usepackage{epstopdf}
\usepackage{epsfig}
\usepackage{amsmath}
\usepackage{mhchem}
\usepackage{units}

\begin{document}

\title{Super-resolution microscopy of single rare-earth emitters}

\author{R. Kolesov$^1$, S. Lasse$^1$, C. Rothfuchs$^2$, A.D. Wieck$^2$, K. Xia$^3$, T. Kornher$^1$, and J. Wrachtrup$^1$\\
$^1$3. Physikalisches Institut, Universit\"at Stuttgart, Pfaffenwaldring 57, Stuttgart, D-70569, Germany\\
$^2$ Ruhr-Universit\"at Bochum, Universit\"atsstra\ss e 150 Geb\"aude NB, D-44780 Bochum, Germany \\
$^3$Department of Physics, The Chinese University of Hong Kong, Shatin, New Territories, Hong Kong, China\\
}

\date{\today}

\begin{abstract}
We demonstrate super-resolution imaging of single rare-earth emitting centers, namely, trivalent cerium, in yttrium aluminum garnet (YAG) crystals by means of stimulated emission depletion (STED) microscopy. The achieved all-optical resolution is $\approx$ \unit[80]{nm}. Similar results were obtained on H3 color centers in diamond with resolution of $\approx$ \unit[60]{nm}. In both cases, STED resolution is improving slower than the inverse square-root of the depletion beam intensity. This is
caused by excited state absorption (ESA) and interaction of the emitter with non-fluorescing crystal defects in its near surrounding.
\end{abstract}

\maketitle

Super-resolution imaging techniques, such as STED \cite{STED}, STORM \cite{STORM}, etc. revolutionized the field of microscopy especially with respect to its bio-applications. They are also having large potential in all-optical addressing of closely spaced quantum systems for quantum information processing protocols. So far, super-resolution imaging is routinely performed with a variety of fluorescent dyes as markers. The techniques mentioned above allow for resolving individual fluorescent molecules spaced much closer than allowed by the diffraction limit. The main disadvantage of these emitting species is their short lifetime due to photobleaching. The only photostable species imaged with super-resolution as single emitting centers was nitrogen-vacancy (NV) center in diamond \cite{NV-STED}. NV centers in diamond allowed for the finest resolution of all-optical imaging of fluorescent species by using STED microscopy \cite{NV-STED-best}. In relation to quantum applications, it is important that STED does not destroy spin properties of the emitting center. This gives an opportunity of optical addressing two nanometer-spaced spin qubits.

\begin{figure}
\center{
\includegraphics[width=0.48\textwidth]{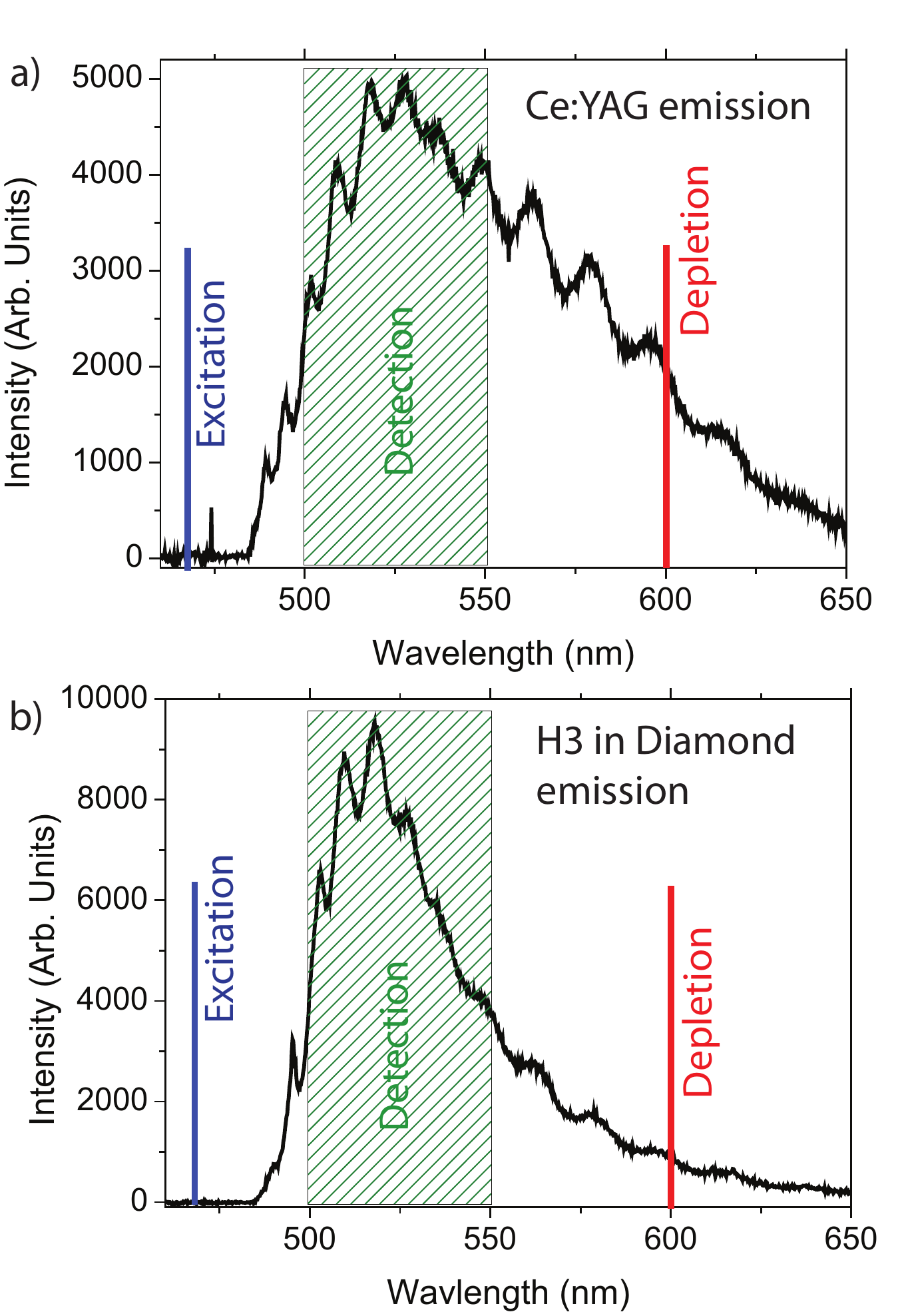}
\caption{\label{fig:Spectra} Emission spectra of a) \ce{Ce^3+} ion in YAG crystal and b) H3 center in diamond (single emitters). Both species can be efficiently excited by \unit[470]{nm} radiation and depleted by \unit[600]{nm} laser. In both cases, quite narrow detection window was used in order to efficiently reject the depletion laser. Slight spectral wiggles are introduced by interference in the optical path and are unphysical.}
}
\end{figure}

In this work, we report on STED microscopy of individual \ce{Ce^3+} centers in YAG single crystals. Since cerium electron spin can be used as a qubit \cite{Ce-qubit}, this paves a way to all-optical addressing of these qubits. As a side result of this study, we also report on STED microscopy of photostable H3 centers in diamond since their excitation and emission spectra are very similar to the ones of cerium in YAG (see Fig.\ref{fig:Spectra}).

Excitation and emission spectra of \ce{Ce^3+} in YAG are studied very well \cite{CeYAG-spectra}. Cerium fluorescence due to $5d\rightarrow 4f$ electronic transition can be excited with blue light with the maximum efficiency at around \unit[460]{nm} and detected in the green-orange range with the maximum at \unit[550]{nm}. Since the red tail of cerium emission extends far into the red up to \unit[700]{nm}, it can be used to stimulatedly deplete the excited state population.

\begin{figure}
\center{
\includegraphics[width=8cm]{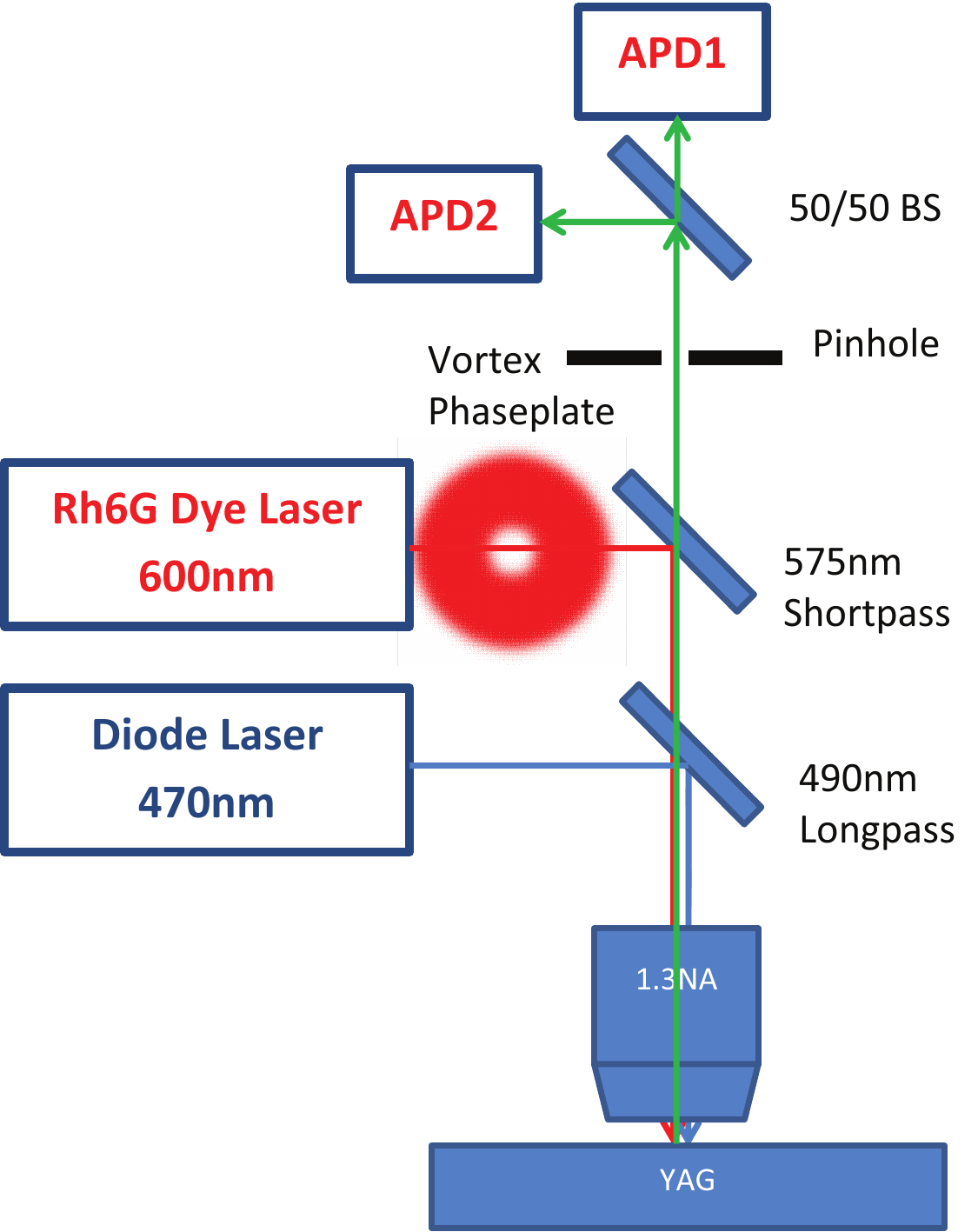}
\caption{\label{fig:setup} Schematic diagram of the STED microscope.}
}
\end{figure}

All experiments presented in this work were performed at room temperature in a home-built STED microscope (see Fig.\ref{fig:setup}). To excite cerium fluorescence, we used a laser diode operating at \unit[470]{nm} wavelength. The excited state population was depleted by a dye laser operating close to \unit[600]{nm} corresponding to the red tail of \ce{Ce^3+} fluorescence. Both excitation and depletion lasers were passed through single mode polarization maintaining optical fibers to obtain Gaussian profile of the beam, recombined on the dichroic beamsplitter, and sent onto the oil-immersion microscope objective lens (Zeiss Fluar, 100x1.3NA). In order to form donut-shaped depletion beam, a vortex plate of charge 1 from RPC Photonics was installed in the depletion path. The fluorescence signal was split into two paths and detected with two single photon detectors forming a Hanbury-Brown and Twiss correlation setup. This allowed us to confirm that the fluorescent centers are indeed single quantum emitters. The spectral detection window was squeezed between \unit[500]{nm} and \unit[550]{nm} by interference filters in order to reject light from both excitation and depletion lasers.

Ultrapure YAG crystal used in this study was purchased at Scientific Materials Corp. In our previous studies, we have already shown that cerium can be stabilized in its trivalent state by high temperature annealing in hydrogen-argon atmosphere \cite{Ce-implant}. After annealing for 24 hours at \unit[1200]{$^{\circ}$C} in an atmosphere of 5\% \ce{H2}/95\% \ce{Ar} at $\sim$ \unit[1]{mbar} total pressure, natural cerium impurities in the crystal showed stable emission under continuous wave (CW) excitation with \unit[470]{nm} laser diode without blinking or bleaching. In order to increase STED resolution, the power of excitation laser was kept below the saturation threshold of \ce{Ce^3+} ions while the power of the depletion beam was maximized. The latter was up to \unit[60]{mW}
right in front of the back aperture of the objective lens. The two laser beams (excitation and depletion) were carefully overlapped so that they are focused by the objective lens in the same plane and the center of the donut coincides with the maximum of the \unit[470]{nm} beam in the focal plane. In this way, the population of the $5d$ excited state of cerium is efficiently depleted everywhere else but the very center of the excitation beam thus dramatically increasing the imaging resolution. The sample was mounted on a nanopositioning stage and scanned through the focus of the laser beams.

\begin{figure}
\center{
\includegraphics[width=0.48\textwidth]{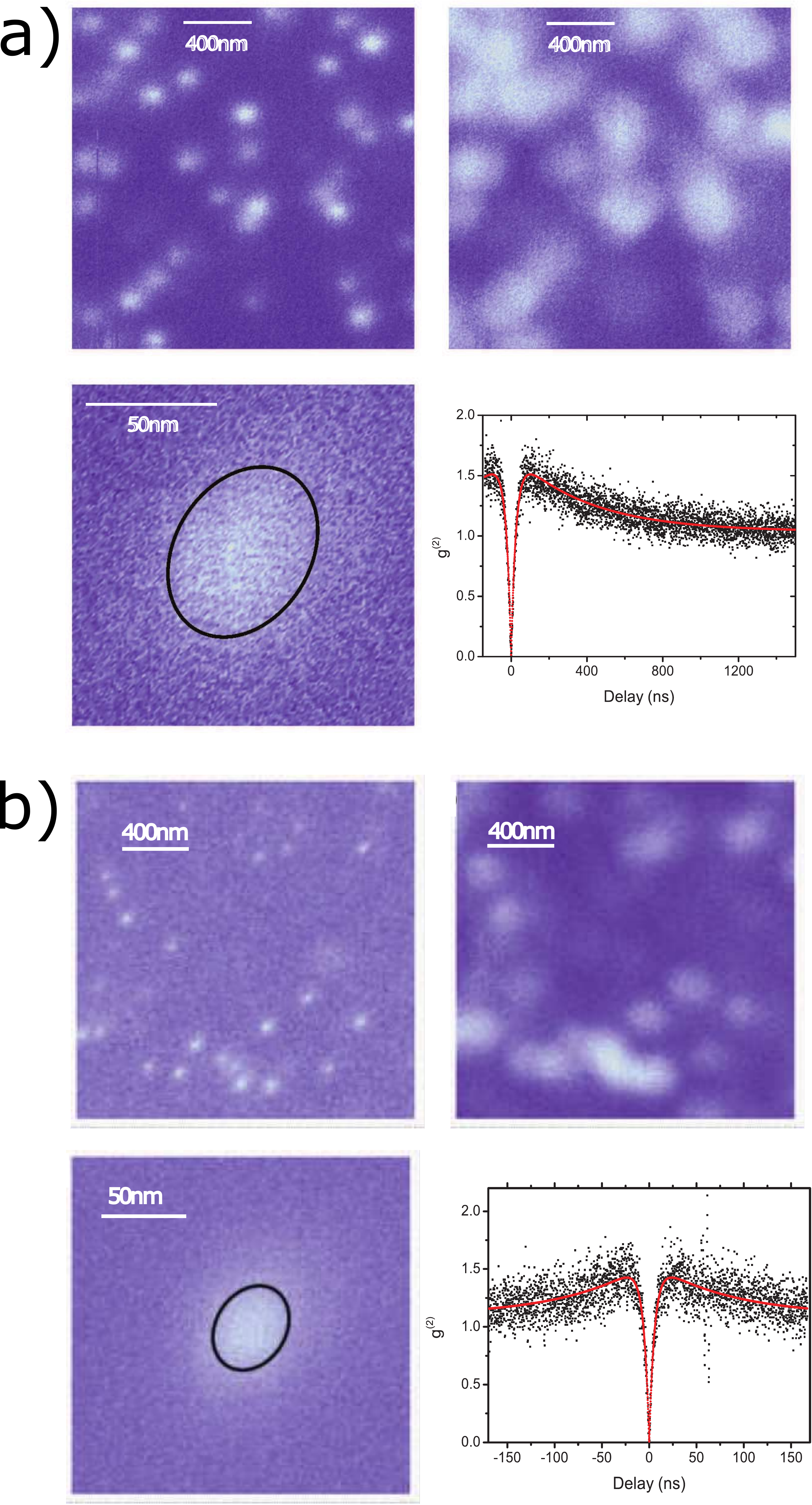}
\caption{\label{fig:Ce-H3-STED}a) Upper left: Super-resolution image of individual \ce{Ce^3+} centers in YAG. Upper right: Image of the same centers obtained with confocal microscope. Lower left: Zoomed image of a single \ce{Ce^3+} center representing PSF of our STED microscope. An ellipsoid shows the section of PSF at its half-maximum. The axes of the ellipsoid are \unit[51]{nm} and \unit[68]{nm}. Lower right: Photon anti-bunching shows that the observed \ce{Ce^3+} center is indeed single. The experimental photon correlation data were corrected for backround according to standard correction procedure \cite{AB-corr}.
b) Upper left: Super-resolution image of H3 centers in diamond. Upper right: Image of the same centers obtained with confocal microscope. Lower left: Zoomed image of a single H3 center representing the PSF of our STED microscope. An ellipsoid shows the section of PSF at its half-maximum. The axes of the ellipsoid are \unit[41]{nm} and \unit[53]{nm}. Lower right: Second order photon correlation function shows that the observed H3 center is indeed single.}
}
\end{figure}

The results of super-resolution imaging of \ce{Ce^3+} ions in YAG are shown in Fig.\ref{fig:Ce-H3-STED}a. In order to assess the improvement in resolution, Fig.\ref{fig:Ce-H3-STED}a presents the confocal scan of the same region of the crystal with the donut depletion beam blocked. One can see the dramatic increase in the resolving power of the STED microscope compared to the confocal case. The achieved resolution was assessed by measuring the point-spread function (PSF) of our STED microscope. The latter was obtained by imaging a single \ce{Ce^3+} center (see Fig.\ref{fig:Ce-H3-STED}a. The PSF was slightly elliptical due to alignment imperfections. Two-dimensional Lorentzian fit revealed the full widths at half maxima of the PSF to be \unit[51]{nm} and \unit[68]{nm}. Thus, the obtained STED resolution can be estimated as \unit[50]{nm}. Fig.\ref{fig:Ce-H3-STED}a shows the results of photon correlation measurements taken on the super-resolution imaged cerium center. The fact that the anti-bunching dip goes below 0.5 indicates that the observed emitter is indeed a single cerium ion.

Similar results were obtained using single H3 centers in diamond. Their spectroscopic properties are very similar to those of \ce{Ce^3+} ions in YAG \cite{H3-spectra}, i.e. they can be excited and depleted with the same wavelengths. The results of STED microscopy of H3 centers are shown in Fig.\ref{fig:Ce-H3-STED}b.

\begin{figure}[h!]
\centering
\includegraphics[width=0.48\textwidth]{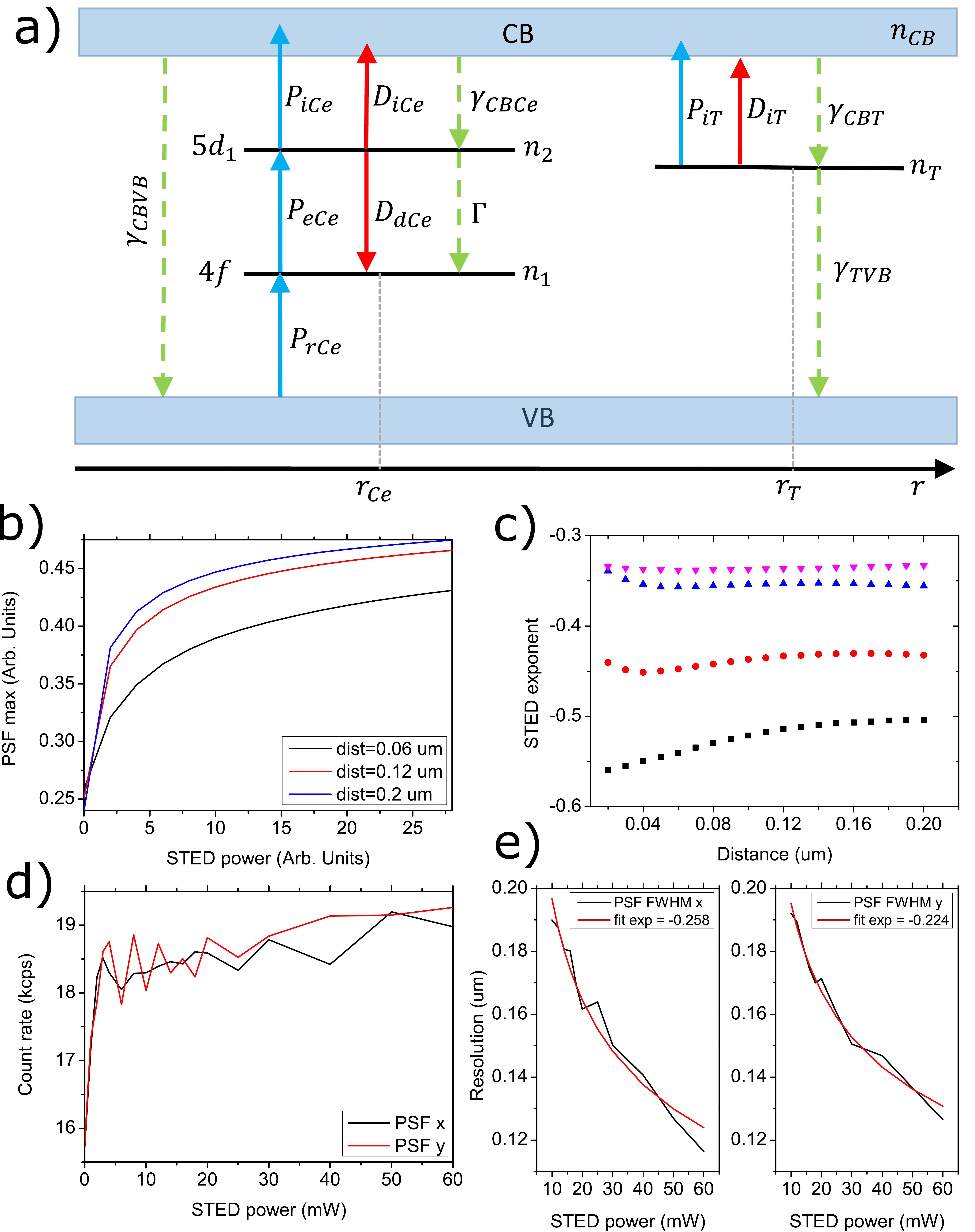}
\caption{a) Diagram of \ce{Ce^3+} levels in YAG accompanied by possible electronic transitions. $P$ stands for pumping at \unit[470]{nm}, {\bf $\Gamma$} stands for fluorescence, $\gamma$ for non-radiative decay and $D$ for depletion donut beam. The indices to $P$ and $D$ correspond to the appropriated cross-sections used. b) Simulated behaviour of the single emitters fluorescence, which was taken to be the maximum value of the PSF for three different distances from the single emitter to the EET. The global simulation parameters were $\Gamma = 1$, $\gamma_{CBCe}=400$, $\gamma_{CBT}=400$, $\gamma_{CBVB}=0.1$. c) Simulated STED exponents for different distances between cerium and EET and for different $\gamma_{TVB}$: 0.04 (purple), 0.2 (blue), 1 (red) and 5 (black).  The global simulation parameters were the same as in b). d) Measured PSF maximum vs. STED power of a single cerium. e) Measured FWHM of PSF vs. STED power (black) and power law fit giving the exponential (red).}
\label{fig:simPSFmaxVSstedPower}
\end{figure}

\begin{figure}[h!]
\center{
\includegraphics[width=10cm]{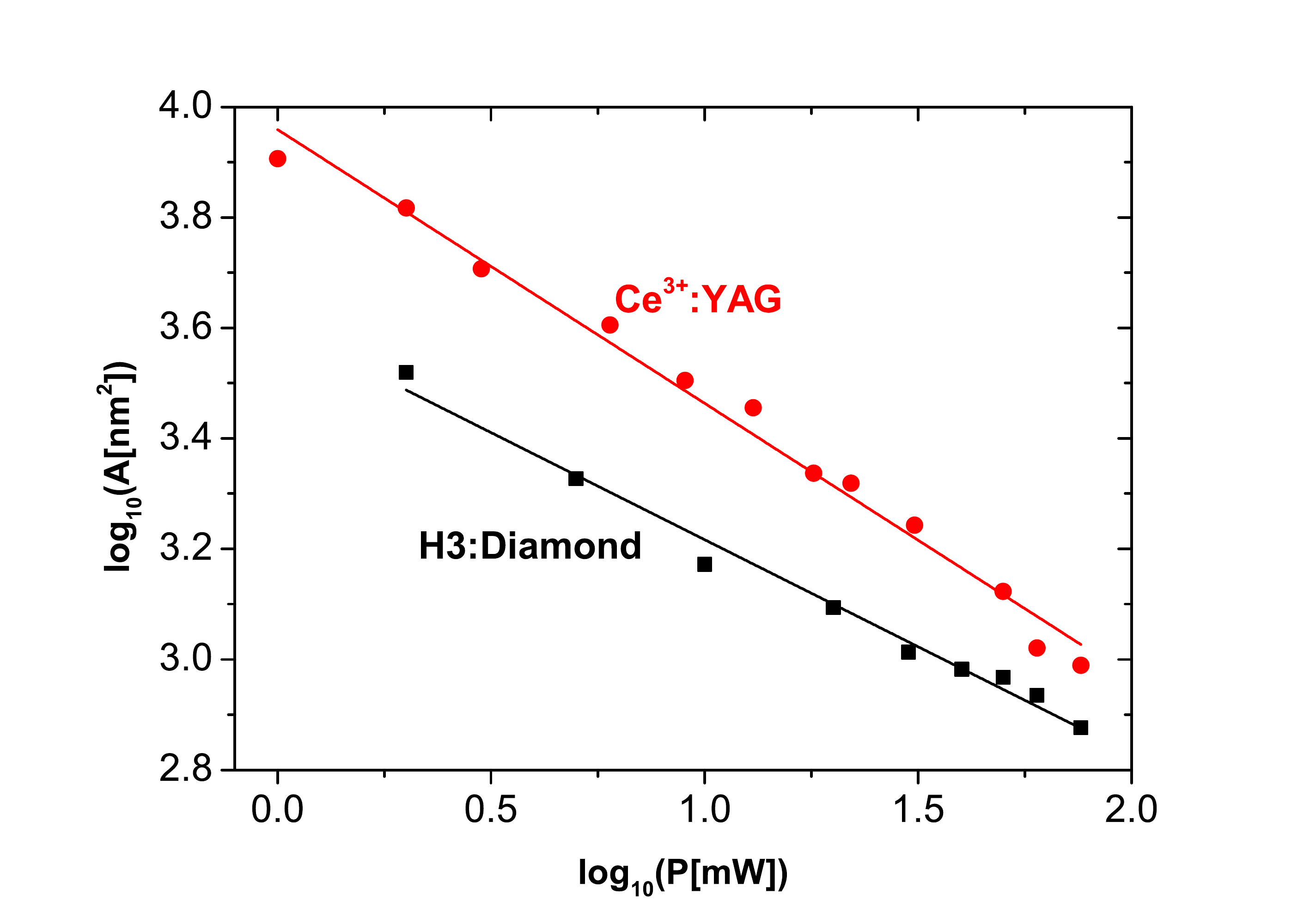}
\caption{\label{fig:resolution} Double logarithmic plot of the area of super-resolution PSF versus depetion beam power. The dependencies are $A\propto P^{-0.5}$ in case of \ce{Ce^3+}:YAG and $A\propto P^{-0.4}$ in case of H3 center in diamond.}
}
\end{figure}

The resolution of STED imaging of both \ce{Ce^3+} in YAG and H3 centers in diamond was studied as a function of the depletion beam power. The results are plotted in Fig.\ref{fig:resolution}. They are significantly deviating from the dependence $A\propto P$ expected for pure STED case when the population of the fluorescent state is being reduced only by stimulation with the depletion beam. This significant difference is most probably related to the excited state absorption (ESA) from the fluorescent state into the conduction band. ESA is known to be prohibiting laser action in \ce{Ce^3+}:YAG crystals implying that the ESA cross-section is greater than that of the stimulated emission. \cite{CeYAG_ESA}. It is also known, that \ce{Ce^3+}:YAG crystals exhibit strong thermoluminescence \cite{CeYAG_thermoluminescence} indicating the presence of deep electron traps below the conduction band of YAG.
Figure \ref{fig:simPSFmaxVSstedPower}a proposes the electronic level structure of the system
causing the deviation from the inverse square-root dependence of STED. It includes a single \ce{Ce^3+}, the effective electron trap (EET) slightly delocalized from the \ce{Ce^3+}, the conduction band and the valence band of the YAG crystal. Both the cerium and the EET exhibit ESA with pump and STED laser, which are capable to ionize the electron of both species to the conduction band. From there, the electron can recombine either to the cerium, to the EET or to the valence band. The transitions induced by the pump laser are depicted as blue arrows, those induced by the depletion beam with red arrows. All possible decays are marked with green dashed arrows. As the depletion beam has non-zero intensity at EET position, it can ionize the trap. This increases the probability for the electron to decay to the cerium, resulting in an increased fluorescence even if the cerium is in the center of the donut-shaped beam. 

The  dynamical behaviour of the system and the population distribution of the involved levels are described by a set of balance equations (\ref{eqn:simulation_model}). The $n_i$ represent the population of the ground and excited state of \ce{Ce^3+}, the EET and the conduction band of the crystal. $P(r)$ and $D(r)$ represent the power of the pump and depletion laser at the current positions of \ce{Ce^3+} and the EET. The cross-sections for repumping, excitation and ionization for \ce{Ce^3+} are denoted by $\sigma_{rCe}$, $\sigma_{eCe}$ and $\sigma_{iCe}$ and the cross-section for ionizing the electron trap is indicated as $\sigma_{iT}$. The spontaneous decay of \ce{Ce^3+} is marked as $\Gamma$ and all non-radiative decays from conduction band to valence band, the decay from conduction band to \ce{Ce^3+} and to the electron trap and the decay from the electron trap to the valence band are introduced as $\gamma_{CBVB}$, $\gamma_{CBCe}$, $\gamma_{CBT}$ and $\gamma_{TVB}$.
\begin{align}
\dot{n}_1 &= P(r)\sigma_{rCe}(1-n_1-n_2)+(\Gamma+D(r)\sigma_{dCe})n_2\notag\\
          &\phantom{{}=}-P(r)\sigma_{eCe}n1\nonumber\\
\dot{n}_2 &= -(\Gamma+D(r)\sigma_{dCe})n_2+P(r)\sigma_{eCe}n_1\notag\\
          &\phantom{{}=}-(P(r)\sigma_{iCe}+D(r)\sigma_{iCe})n_2\notag\\
          &\phantom{{}=}+\gamma_{CBCe}n_{CB}(1-n_1-n_2)\label{eqn:simulation_model}\\
\dot{n}_T &= -\gamma_{CBT}n_{CB}(1-n_T)-(P(r)\sigma_{iT}+D(r)\sigma_{iT})n_T\notag\\
		  &\phantom{{}=}-\gamma_{TVB}n_T\nonumber\\
\dot{n}_{CB} &= (P(r)\sigma_{iCe}+D(r)\sigma_{iCe})n_2+(P(r)\sigma_{iT}+D(r)\sigma_{iT})n_T\notag\\
			 &\phantom{{}=}-\gamma_{Ce}n_{CB}(1-n_1-n_2)-\gamma_{T}n_{CB}(1-n_T)\notag\\
			 &\phantom{{}=}-\gamma_{CBVB}n_{CB}\nonumber
\end{align}
Solving the equations (\ref{eqn:simulation_model}) in steady state for $n_2$ under continuous variation of the distance from \ce{Ce^3+} to the center of the donut-shaped beam, to the electron trap and for different STED beam intensities results in the spatially dependent PSFs of the simulated system. Integrating the PSFs above its half maximum and plotting the result versus STED intensity gives the evolution of the microscope resolution dependent on the STED power. This is found to improve slower than the inverse square-root of the depeletion beam intensity as it was observed in the experiment.

A secondary effect observed was a ~30\% volatile and then quickly saturating increase in the count rate of the cerium when turning on the depletion beam and then increasing its power. The suggested model reproduces this behaviour for when starting with $D(r)=0$, which corresponds to confocal microscopy, and then increasing the amplitude of the depletion beam.

The above qualitative picture suggests that all fluorescent species exhibiting excited state absortion at the wavelength of the STED laser beam should behave similar. Thus, the resolution of STED microscopy on the species with ESA at STED wavelength is always degraded by the diffusion of electrons in the conduction band of the host material.

In conclusion, we have demonstrated super-resolution imaging of individual photostable cerium centers in yttrium aluminum garnet and H3 centers in diamond. In both cases, optical resolution of \unit[50]{nm} or better is achieved. The dependence of STED resolution on the intensity of the depleting beam deviates from the inverse-square-root and suggests strong influence of ESA and electron dynamics with neighbouring electron traps below the conduction band of YAG, which can be treated as a single effective electron trap.

The work was financially supported by Deutsche Forschungsgemeinschaft (DFG Grants KO4999/1-1 and KO4999/3-1). The authors thank Dr. Andrej Denisenko for useful discussions.

\end{document}